\documentclass[aps,
		prd,
		reprint,
		twocolumn,
		superscriptaddress,
		shortbibliography,
		nofootinbib,
		floatfix,
		notitlepage
		]{revtex4-1}
\pdfoutput=1

\usepackage{amsmath,amssymb}
\usepackage{graphicx,accents}
\usepackage{graphicx,epstopdf}
\usepackage[usenames,dvipsnames]{xcolor}
\usepackage{slashed}
\usepackage[normalem]{ulem}
\usepackage{bm}
\usepackage{bbm}
\usepackage{bbold}

\usepackage[printwatermark]{xwatermark}

\usepackage{hyperref}
\hypersetup{colorlinks=true}

\usepackage{tabularx}

\newcommand{\vphi}{\varphi}
\newcommand{\eps}{\varepsilon}

\newcommand{\mbf}[1]{\mathbf{#1}}
\newcommand{\trm}[1]{\textrm{#1}}
\newcommand{\tsf}[1]{\textsf{#1}}

\newcommand{\be}{\begin{equation}}
\newcommand{\ee}{\end{equation}}
\newcommand{\bea}{\begin{eqnarray}}
\newcommand{\eea}{\end{eqnarray}}
\newcommand{\bi}{\begin{itemize}}
\newcommand{\ei}{\end{itemize}}
\newcommand{\sinc}{\textrm{sinc}}

\newcommand{\nn}{\nonumber}

\definecolor{aaaa}{rgb}{0.99, 0.4, 0.01}
\definecolor{bbbb}{rgb}{0.5, 0.3, 0.9}
\definecolor{bk1}{RGB}{230,40,40}

\newcommand{\figref}[1]{Fig. \ref{#1}}
\newcommand{\figrefa}[1]{Fig. \ref{#1}(a)}
\newcommand{\figrefb}[1]{Fig. \ref{#1}(b)}

\newcommand{\eqnref}[1]{Eq. (\ref{#1})}

\newcommand{\vkap}{\varkappa}

\newcommand{\ud}{\mathrm{d}}

\newcommand{\LCperp}{{\scriptscriptstyle \perp}}

\makeatletter
\def\ps@pprintTitle{%
 \let\@oddhead\@empty
 \let\@evenhead\@empty
 \def\@oddfoot{}%
 \let\@evenfoot\@oddfoot}
\makeatother

\begin{document}

\title{Pulse envelope effects in nonlinear Breit-Wheeler pair-creation}
\author{S.~Tang}
\affiliation{College of Physics and Optoelectronic Engineering, Ocean University of China, Qingdao, Shandong, 266100, China}

\author{B.~King}
\email{b.king@plymouth.ac.uk}
\affiliation{Centre for Mathematical Sciences, University of Plymouth, Plymouth, PL4 8AA, United
Kingdom}


\date{\today}
\begin{abstract}
The effect of the pulse envelope on electron-positron pair creation in a circularly-polarised laser pulse is investigated. Interference on the length scale of the pulse envelope, and smoothness of the pulse edges are found to influence the pair spectrum. A toy model of a flat-top pulse is used to identify pulse envelope effects inaccessible to local approaches. Broadening of channel openings and a widening of the energy and transverse momentum distribution of the pair are found to receive contributions that are below the local harmonic threshold. By comparing pair yields in a flat-top, sine-squared and Gaussian pulse, a link between pulse shape and photon-polarised Breit-Wheeler is found. In the transverse momentum distribution, a signal of pulse envelope interference is found in an azimuthal asymmetry, which appears in intense fields and persists in long pulses.
\end{abstract}
\maketitle
\twocolumngrid

\section{Introduction}
The conversion of two photons to an electron-positron pair was calculated by Breit and Wheeler in the 1930s \cite{breit34}. The decay of a \emph{single} photon into an electron-positron pair in a classical electromagnetic field, is often referred to as the `nonlinear Breit-Wheeler' process. `Nonlinear' refers to the dependency on the charge-field coupling, which generally can depend on high powers of the field intensity. One of the challenges in calculating this process in a classical background is that the amplitude can depend strongly on how the background varies in space and time. This can include interference from the process happening at different points in the background, and gradient effects from the rising and falling edge of the pulse. These effects can mean that if the process is approximated by splitting the background into infinitessimally short intervals, using the probability for the process to occur in a constant (and crossed) field in each interval and then integrating over all the intervals the background was split into, the result can differ substantially from the correct one. For high field intensities
this procedure can, however, in general provide a good approximation, which is often referred to as the `locally constant field approximation' (LCFA) \cite{ritus85,harvey15,DiPiazza:2017raw,Ilderton:2018nws,DiPiazza:2018bfu,King:2019igt,Seipt:2020diz}, and is the central approximation by which quantum electrodynamical (QED) effects in intense classical backgrounds, such as the nonlinear Breit-Wheeler process \cite{reiss62,nikishov64,narozhny69,heinzl10,PhysRevA.84.033416,PhysRevA.86.052104,PhysRevLett.108.240406,seipt12b,PhysRevA.88.062110,Titov:2013kya,Meuren:2014uia,Wu:2014zaa,Jansen:2015idl,Meuren:2015mra,Nousch:2015pja,PhysRevLett.117.213201,PhysRevD.94.013010,Hartin:2018sha,Ilderton:2019ceq,Ilderton:2019vot,Titov:2019kdk,Titov:2020taw,mercuri.njp.2021}, are included into numerical simulation frameworks \cite{nerush11,PhysRevSTAB.14.054401,RIDGERS2014273,gonoskov15,Gelfer:2015ora,grismayer16,lobet.prab.2017,WAN2020135120,PhysRevLett.123.174801,Seipt:2020uxv}.

In this paper, we will consider the situation where the background field has two timescales. This is a relevant situation for, e.g. experiments involving a laser pulse, which has the fast timescale of the carrier frequency, and the slow timescale of the pulse envelope. For a strongly-focussed laser pulse, there may be other relevant timescales, but we will model the laser pulse here as a plane wave, which should be a good approximation for a weakly-focussed pulse. We will consider laser intensities that are routinely attained with weak focussing in the lab.

The LCFA depends on the single parameter, $\chi$, the `quantum nonlinearity parameter', which has no explicit reference to any length scale of the field. Therefore, the LCFA can be thought to include interference effects on the sub-wavelength scale. (The length scale over which interference effects play a role, is sometimes referred to as the `formation length' \cite{ritus85}.) The advantage is that the LCFA can be employed when the form of the classical background is not known \emph{a priori}. An alternative approximation, which is useful when the form of the background is close to a plane wave and is known not to change substantially in interaction with e.g. a probe beam, is the locally monochromatic approximation (LMA) \cite{Heinzl:2020ynb}. The LMA treats the fast timescale of the carrier frequency exactly and neglects derivatives of the pulse envelope. It depends on two parameters, the classical nonlinearity parameter $\xi$, which is proportional to $\lambda$, the wavelength of the background, and it depends on the (quantum) energy parameter, $\eta$, proportional to $1/\lambda$. Therefore the LMA can be thought to include interference effects on the length scale of the wavelength (the `fast' timescale). In the limit of large $\xi$ and large outgoing particle energy, the LMA depends only on the product $\chi = \xi \eta$, and tends to the LCFA, with all explicit dependency on the background having disappeared.

Motivation for the current study comes from planned upcoming experiments, such as E320 at SLAC and LUXE \cite{Abramowicz:2021zja} at DESY, which will use a conventionally-accelerated electron beam of $O(10\trm{GeV})$, to measure the nonlinear Breit-Wheeler process in the `all-order' region where the charge-field coupling, described by the classical nonlinearity parameter, $\xi$, is $\sim O(1)$. The interaction point of the LUXE experiment is modelled using the simulation code Ptarmigan \cite{ptarmigan}, which implements the LMA for the nonlinear Compton \cite{Blackburn:2021rqm} and nonlinear Breit-Wheeler \cite{Blackburn:2021cuq} processes. (A similar approximation framework \cite{Bamber:1999zt} was implemented to model the E144 experiment \cite{E144:1996enr,burke97}, and also been used in CAIN \cite{cain1} and IPStrong \cite{Hartin:2018egj}.) To understand what is missed in local approaches is therefore worthy of investigation.

The current paper will find two main effects that are missed by local or `instantaneous' approaches. One effect is due to interference on the length-scale of the pulse envelope (i.e. the `slow' timescale). Therefore, in addition to $\xi$ and $\eta$, the number of laser cycles, $N$, which is proportional to the pulse length, will be used to specify the input parameters. These effects persist in magnitude (although they may move to lower energies) as the pulse length is increased. Another effect is due to derivatives of the pulse envelope, which are missed by local approaches that only include the leading order derivative (i.e. the gradient) of the background. We will find that these effects can be increased (decreased) in amplitude by making the pulse envelope edges steeper (shallower), as quantified by the bandwidth of the Fourier transform.

To investigate pulse envelope effects, we will use a combination of direct numerical evaluation of QED calculations for a plane-wave pulse as well as an analytical `toy model' of a circularly-polarised `flat-top' monochromatic pulse. This toy model has the advantage that any deviation from the locally-monochromatic spectrum can only be caused by pulse-envelope interference, as it is only at the beginning and end of the pulse where there is a variation in the field strength. Therefore these deviations are \emph{beyond} a locally monochromatic (and therefore beyond a locally constant) approximation.
\newline

One may think that interference effects over long spacetime scales are less important for pair-creation than e.g. for Compton scattering because pair-creation is typically strongly suppressed for lower field strengths, and so only a small region around the laser pulse peak field strength would contribute. (This is an explanation that follows when $a_{0}\gg 1$, see e.g. \cite{Dinu:2019wdw}.) However, for lower field strengths, it is exactly this strong suppression, which makes the process sensitive to bandwidth effects around the carrier frequency of the pulse.

The effect the pulse envelope can have on nonlinear Breit-Wheeler has been investigated already in several works by direct numerical evaluation of the QED expressions, or by using a `slowly-varying' envelope approach. Line-broadening and sub-threshold enhancement have been identified in triple differential spectra \cite{heinzl10}, and an enhancement in energy spectra in the multi-photon region in circularly- \cite{PhysRevLett.108.240406} and linearly-polarised \cite{seipt12b} laser pulses has been linked to finite bandwidth effects.  Short, circularly-polarised pulses lead to an asymmetry in the azimuthal spectrum of produced pairs \cite{Seipt:2013taa}, which is sensitive to the pulse's carrier envelope phase \cite{Krajewska:2012eb,Titov:2015tdz}. When double pulses are used, various pulse shape effects can arise and be controlled \cite{Nousch:2015pja,Jansen:2016crq,PhysRevD.98.036022,Peng:2018hmj}, and multiple pulses can lead to coherent enhancement \cite{Krajewska:2014ssa}. There also exist some analytical solutions for nonlinear Breit-Wheeler in special ultra-short plane-wave pulses \cite{Ilderton:2019ceq,Ilderton:2019vot}. Apart from Breit-Wheeler pair creation, the pulse envelope's shape can also be crucial in determining pair creation via the dynamically-assisted Sauter-Schwinger effect \cite{Schutzhold:2008pz,Kohlfurst:2012rb,Gies:2015hia,Torgrimsson:2017pzs,Huang:2019uhf}.

The paper is organised as follows. In Sec. II the flat-top potential toy model is introduced and the effect on the total yield of pairs, the lightfront momentum spectrum, and the angular spectrum is illustrated. In Sec. III, a leading-order perturbative analysis is given, corresponding to linear Breit Wheeler, and a link between pulse duration and photon polarisation shown. In Sec. IV, the direct numerical evaluation of the QED expressions for a flat top, sine-squared and Gaussian pulse are presented and azimuthal asymmetry in the angular spectrum of pairs is identified as a signal of pulse interference. We use natural units $\hbar=c=1$ throughout and the fine structure constant is $\alpha=e^2\approx1/137$.

\section{Toy model: flat-top potential}
Consider a circularly polarised potential of the form:
\[
a = m\xi (0, \cos \vphi, \sin\vphi,0 ),\quad 0<\vphi<\Phi,
\]
otherwise $a = 0$, where $a=|e|A$ and $\vphi=\vkap\cdot x$ is the background phase, and $\vkap = \vkap^{0}(1,0,0,1)$, where $\vkap^{0}$ is the laser carrier frequency. The pulse phase length is $\Phi = 2N\pi$, where $N$ is the number of the laser cycles. The energy parameter, $\eta=\vkap\cdot k/m^{2}$, is related to the centre-of-mass energy of the collision. In \cite{King:2020hsk}, nonlinear Compton scattering was studied in this background, here we calculate nonlinear Breit-Wheeler.

Here, we give the main expressions and definitions for the probability of nonlinear Breit-Wheeler, $\tsf{P}$ (more details about the method can be found in~\cite{TangPRA2021}).
\begin{align}
\tsf{P}=&\frac{\alpha}{(2\pi\eta)^2} \int \frac{d t\, d^{2}\mbf{r}^{\LCperp}}{t(1-t)}\int d\vphi d\vphi'~ e^{i\int^{\vphi}_{\vphi'}d\phi\frac{k\cdot \pi_{q}(\phi)}{\eta m^2(1-t)}}\nonumber\\
&\left[\Delta \Delta'+h(t)\left(a^{2}\Delta'-2a\cdot a'+a'^{2}\Delta \right)/(2m^2)\right]\,,
\label{eqn:P1}
\end{align}
where $t=\vkap\cdot q /\vkap\cdot k$, $q^{\mu}$ is the asymptotic momentum of the positron after leaving the potential and $k^{\mu}$ is the photon momentum, $\mbf{r}^{\LCperp} = \mbf{q}^{\LCperp}/m - t\mbf{k}^{\LCperp}/m$, $a=a(\vphi)$ and $a'=a(\vphi')$, with  $h(t)=(1-2t+2t^{2})/[2t(1-t)]$.  The instantaneous momentum of the positron in the field, $\pi_{q}$ is:
\bea
 \pi_{q} = q - a + \frac{q\cdot a}{ k\cdot q}\vkap-\frac{a^2}{2k\cdot q}\vkap\,,
\label{eqn:POSmom}
\eea
and
\bea
\Delta = 1 - \frac{k\cdot \pi_{q}}{k\cdot q} = \frac{2\, m\mbf{r}^{\LCperp}\cdot \mbf{a}-\mbf{a}^{2}}{m^{2}[1+(\mbf{r}^{\LCperp})^{2}]},\label{eqn:DELfac}
\eea
$\Delta'$ is analogous to $\Delta$ but with $\mbf{a}(\vphi)$ replaced with $\mbf{a}(\vphi')$.

To specialise \eqnref{eqn:P1} to the flat-top potential, we employ the Jacobi-Anger expansion and integrate over the azimuthal transverse 
co-ordinate. This yields a total probability that can be written as a sum over harmonics, $n$: $\tsf{P} = \alpha /\eta\sum_{n= \lceil n_{\ast} \rceil}^{\infty} \mathcal{I}_{n}$~ where $\lceil n_{\ast} \rceil$ denotes the lowest integer greater than or equal to $n_{\ast}$, and here for the flat-top potential, $n_{\ast}=-\infty$. Then $\mathcal{I}_{n}$ is given by\footnote{The result in \eqnref{eqn:Inflat} has been numerically verified by direct evaluation of \eqnref{eqn:P1} in a flat-top background.}:
\bea
\mathcal{I}_{n} &=& \frac{\Phi}{\eta}\int \frac{dt\, dr^{2}}{t(1-t)} \, \delta_{\Phi}\left[\frac{r^{2}-r_{\infty}^{2}}{2\eta t(1-t)}\right] \left\{ w^{2}J_{n}^{2}(z)  \right. \nn \\
&& \left. - \xi^{2}h(t)\left[2wJ_{n}^{2}(z)-J_{n+1}^{2}(z)-J_{n-1}^{2}(z)\right]/2\right\}, \label{eqn:Inflat}
\eea
with $r = |\mbf{r}^{\perp}|$,
where $z=\xi r / \eta t(1-t)$ and we have defined the regularised delta function
\bea
\delta_{\Phi}(x)=\frac{\Phi}{2\pi}~\trm{sinc}^{2}\left(\frac{\Phi x}{2}\right),
\eea
($\trm{sinc}\,x = \sin x/x$) and the finite-duration factor $w$ is:
\[
w = \frac{1+r_{\infty}^{2}}{1+r^{2}};\qquad r_{\infty}^{2}=2n\eta t(1-t) - (1+\xi^{2}).
\]
The function $\delta_{\Phi}(x)$ tends to the Dirac delta function $\delta(x)$ in the infinite pulse-length limit of $\Phi\to\infty$. In this limit, the finite-duration factor $w\to1$ and \mbox{Eq.~(\ref{eqn:Inflat})} tends to the locally monochromatic (and infinite monochromatic \cite{nikishov64}) result. As remarked in \cite{King:2020hsk}, the function $\delta_{\Phi}$ can be written in terms of harmonics as:
\bea
\delta_{\Phi}(n-\tilde{n}_{\ast}),\qquad \trm{where: }\tilde{n}_{\ast} = \frac{1+\xi^{2}+r^{2}}{2\eta t(1-t)}. \label{eqn:deltaN}
\eea
Our analysis of the result, \eqnref{eqn:Inflat}, will focus on the description in terms of harmonics.

\subsection{Harmonic channel opening}
In the locally monochromatic approach, we can acquire a `threshold' harmonic $n_{\ast} \to\tilde{n}_{\ast}$ from~(\ref{eqn:deltaN}), by finding the minimum of $\tilde{n}_{\ast}$, which occurs when $r=0$ and $t=1/2$, i.e. \mbox{$n_{\ast}=2(1+\xi^{2})/\eta$}. When the locally monochromatic approach is applied to a pulse, $a(\vphi)$, different parts of the pulse can have different threshold harmonics, and therefore access different harmonic `channels' to pair creation. In the local approach, we can see that having a non-constant pulse envelope implies that some channels become accessible, that would otherwise not be, if just the average pulse intensity was assumed. However, in the flat-top model, because $a(\vphi)$ is constant everywhere (apart from the edges of the pulse), this local effect \emph{cannot} access any channel opening behaviour.

When we take into account the finite pulse effect and evaluate \eqnref{eqn:Inflat}, we see several differences to the local approach.
\bi
\item Each harmonic is `broadened'. In the local approach, as $\delta_{\Phi}(\cdot)\to\delta(\cdot)$ once the momentum of the pair is fixed, so is the harmonic order $n=\tilde{n}_{\ast}$. However, since the pulse is finite in phase, it has a finite bandwidth in the lightfront momentum it can supply to the pair.
\item There is no longer a threshold harmonic. The central peak of $\delta_{\Phi}$ has a width such that the main contribution is for a harmonic $n$ within the range approximately bounded by $\tilde{n}_{\ast} \pm \pi/\Phi$. The $n=0$ channel is opened and short pulses allow for the contribution from negative harmonics.
\item The wavevector of the background $\vkap$, and therefore also the energy parameter $\eta$ is no longer a constant. A Fourier transform of the background of a pulse reveals a width of momenta around the ``central'' carrier envelope frequency.
\ei

\begin{figure}[h!!]
    \centering
        \includegraphics[width=6cm]{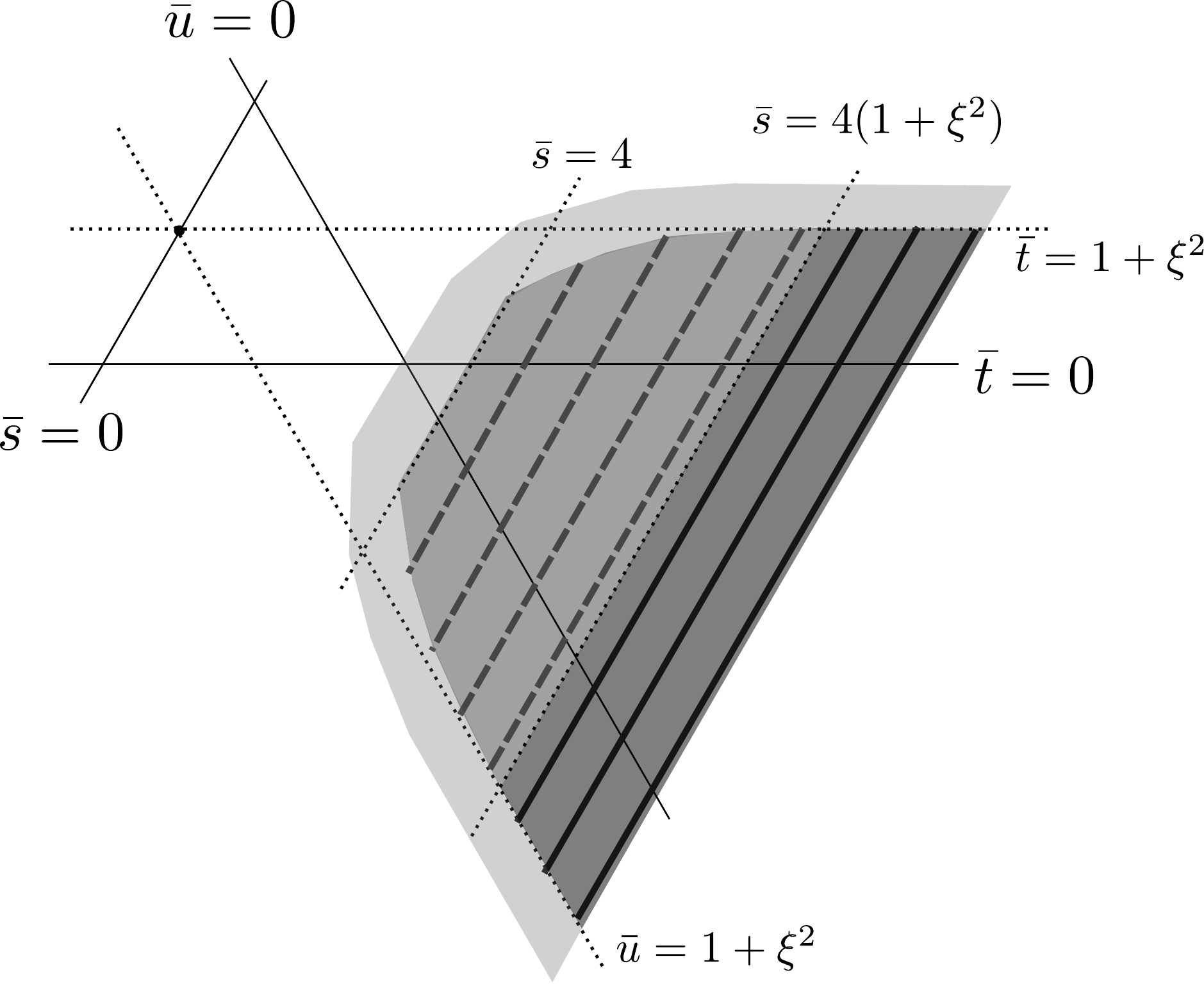}
    \caption{Mandelstam plot for the Breit-Wheeler process. Kinematically accessible regions are highlighted: Monochromatic case (darkest shaded region, solid harmonics lines); locally monochromatic (mid-darkest shaded region, long-dashed harmonic lines); finite pulse (bandwidth suggested by lightest shaded region).}
    \label{fig:mandelstam1}
\end{figure}

The situation can be demonstrated by the use of a Mandelstam plot. Using momentum conservation in a monochromatic wave
\bea
k + n \vkap = \bar{p} + \bar{q}, \label{eqn:mom1}
\eea
where $\bar{p} = p - \vkap (a^{2}/2\vkap \cdot p)$ (and likewise for $\bar{q}$) is the quasimomentum of the electron (positron), we define the Mandelstam variables: $\bar{s}=(k+n\vkap)^2/m^{2}=2 n \eta$, $\bar{t}=(\bar{q}-n\vkap)^2/m^{2}=1+\xi^{2}-2t n\eta$ and $\bar{u}=(\bar{p}-n\vkap)^2/m^{2}=1+\xi^{2}-2(1-t) n\eta$. We see that, for fixed background momentum $\vkap$ and positive harmonic $n$, the variables $\bar{t}$ and $\bar{u}$ cannot exceed $1+\xi^{2}$.  This defines a physically accessible region, highlighted in \figref{fig:mandelstam1}.  For the monochromatic case, $\xi$ is constant and using the threshold condition $n_{\ast}>2(1+\xi^{2})/\eta$, we see that $\bar{s}\geq 4(1+\xi^{2})$. This region is highlighted in  \figref{fig:mandelstam1}, where for each harmonic, there is a curve of values in the $(\bar{t},\bar{u})$ plane given by varying the parameter $t$ between $0$ and $1$. The locally-monochromatic region would in general allow access to `subthreshold' harmonics because $0\leq\xi(\vphi)\leq\xi$. The corresponding region is also highlighted in the figure.
In contrast, in a finite pulse, \eqnref{eqn:mom1} no longer holds absolutely, due to the pulse edges spoiling the periodic symmetry along the pulse propagation direction. This non-conservation is quantified by the regularised delta function \eqnref{eqn:deltaN}. In principle the bandwidth can be wide enough that all harmonics, including $n=0$ and negative $n$, are kinematically accessible. (In the figure, the finite pulse region is suggested as stretching into a region around the one accessible by a local approach).    
This is particularly relevant for a flat-top pulse, the Fourier transform of which is a $\sinc$-squared function, which, having an inverse-square dependency on frequency, has a particularly wide bandwidth (this point will be developed later).

\begin{figure}[h!!]
    \centering
    \includegraphics[width=6.0cm]{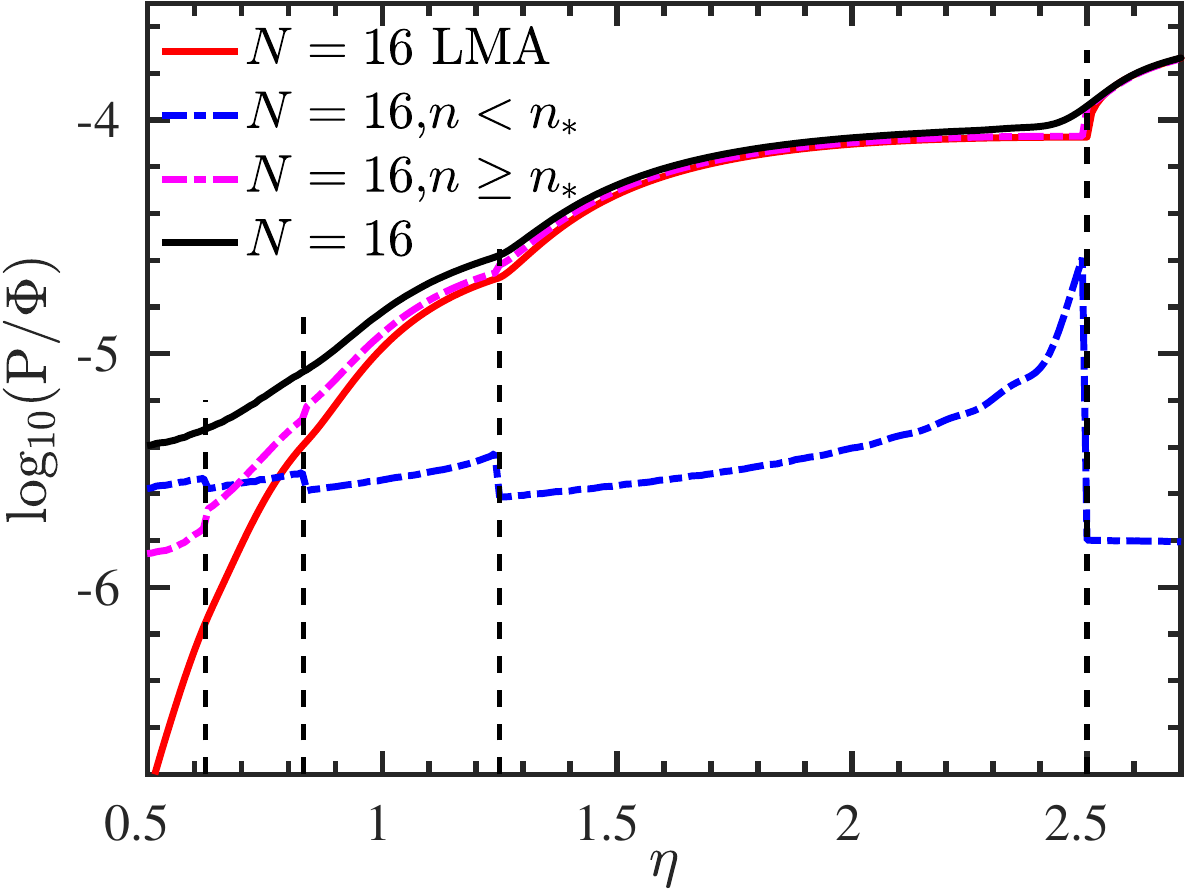}
    \caption{The opening of harmonic channels as the collision energy parameter $\eta$ is increased, for $\xi=0.5$ and $N=16$ laser cycles. Vertical dashed lines correspond to the position of the lowest harmonic in the locally monochromatic approach: $n_{*}=1,2,3,4$ from right to left. The lowest harmonic accessible in the local approach is given as $\lceil n_{\ast} \rceil$, where $n_{\ast}=2(1+\xi^{2})/\eta$.}
    \label{fig:staircase}
\end{figure}

We calculate the effect of these new harmonic channels on the total pair yield in \figref{fig:staircase}. Their contribution can be seen most clearly in two places: i) in the low-energy region $\eta \ll 1$; ii) for parameter values close to a harmonic transition.

Close to a harmonic transition, the `broadness' of the harmonic 
can be clearly seen in \figref{fig:staircase}. Even though the number of laser cycles $N$ (related to pulse duration via $\Phi=2\pi N$) is large enough to be in the long-pulse regime where one would expect the locally-monochromatic approximation to work well, we see that the contribution from the new harmonic channels remains approximately independent of energy, with rising `edge' structures close to each harmonic.

Also in the multiphoton-regime $\xi\ll1$, whereas a local approach gives a pair yield that is power-law suppressed as $\sim\xi^{2n_{\ast}}$ \cite{brezin70}, which for small enough $\xi$, leads to a strong suppression, the flat-top pulse result does not suffer such a suppression. The reason for this is that the finite pulse result has access to the \emph{linear Breit-Wheeler} process, due to the broad bandwidth of the flat-top envelope, which is kinematically forbidden in the local approach.
(This fact has already been investigated in several works in the context of universality \cite{Gies:2015hia,Gies:2016coz}, but here we link it explicitly to a harmonic analysis.)
(The linear Breit-Wheeler case is studied in more detail later on.) Therefore this `enhancement' at small $\eta$ or small $\xi$ is to do with the smoothness of the pulse shape. We note that this enhancement is contributed to by harmonics \emph{above} and \emph{below} threshold.

\subsection{Lightfront momentum spectrum}
\begin{figure}[h!!]
    \centering
    \includegraphics[width=8.5cm]{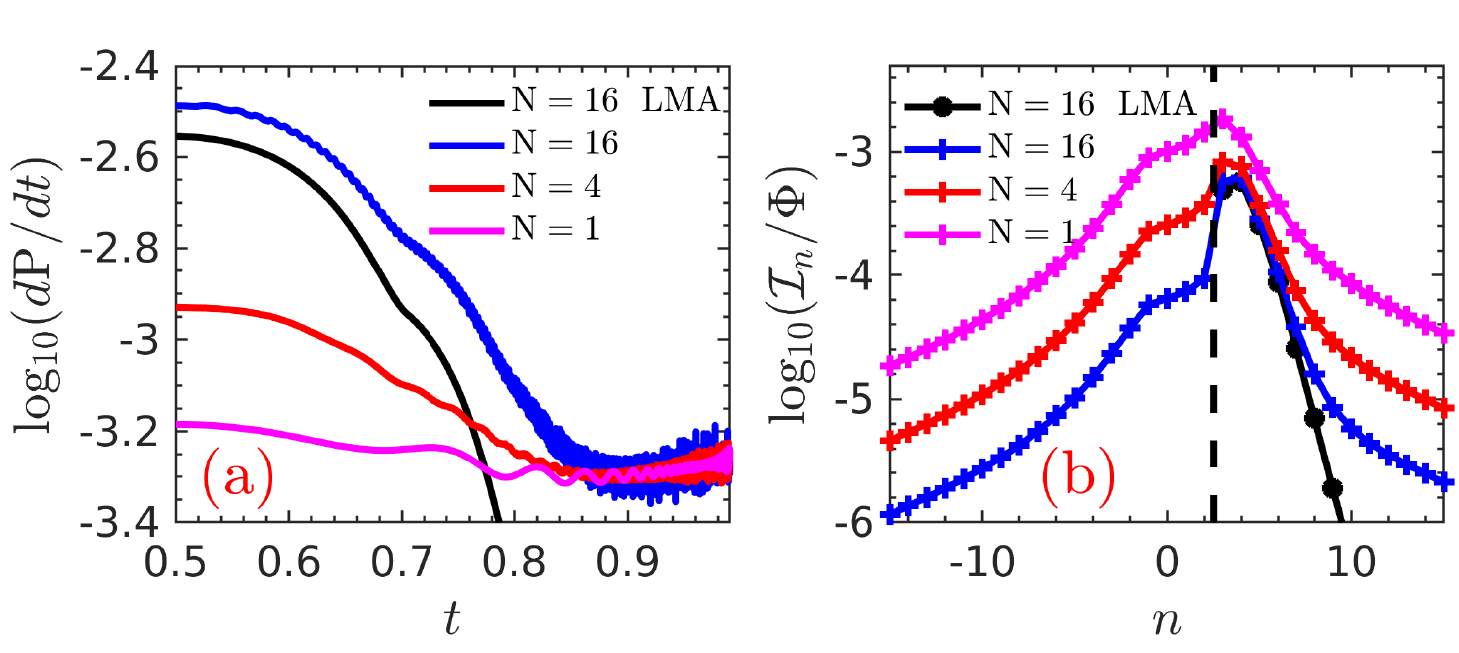}
    \caption{For $\xi=0.5$ and $\eta=1$. Left: the lighfront momentum fraction spectrum of the produced positrons. Because of the symmetry $d\tsf{P}(t)/dt=d\tsf{P}(1-t)/dt$ in~(\ref{eqn:P1}), we only plot the spectrum with $t\geq 0.5$. Right: the contribution of each harmonic, $n$, where the vertical dashed line is the lowest harmonic accessible in the locally-monochromatic case.}
    \label{fig:tplot1}
\end{figure}

The finite bandwidth of the background also affects the shape of lightfront momentum spectra. In the local approach, most pairs are produced with the electron and positron having similar lightfront momenta (i.e. the spectrum has a peak around $t=1/2$.). The probability for one of the pair to take much more of the probe photon's lightfront momentum than the other, is heavily suppressed (i.e. the spectrum is suppressed at large/small $t$). However, in the flat-top case, shorter pulses can greatly enhance the \emph{relative} proportion of pairs whose consitutents can carry very different lightfront momenta. This is illustrated in \figrefa{fig:tplot1}: the shorter the pulse, the flatter the spectrum, but even for long pulses, the enhancement is clearly visible for higher values of asymmetry as $t\to 1$. In particular, the \emph{absolute} enhancement is \emph{independent} of the pulse duration, suggesting it originates in the higher derivatives at the rising and trailing edges of the pulse. (We calculated the lightfront momentum spectrum in the smoother pulse shapes of a sine-squared and Gaussian background, and the asymmetry effect diminishes substantially, showing it to be a consequence of the wide bandwidth of a flat-top pulse.) Which harmonics are contributing to this enhancement, is demonstrated in \figrefb{fig:tplot1}. We see that, indeed the maximum contribution agrees with the locally monochromatic case, but that there is an important region just below the monochromatic threshold, stretching to $n=-1$, which is entirely missed by local approaches. As the pulse is made shorter, the relative contribution increases from these ``below threshold'' harmonics and the lightfront momentum spectrum flattens.

\subsection{Transverse momentum distribution}
\begin{figure}[h!!]
    \centering
    \includegraphics[width=8.5cm]{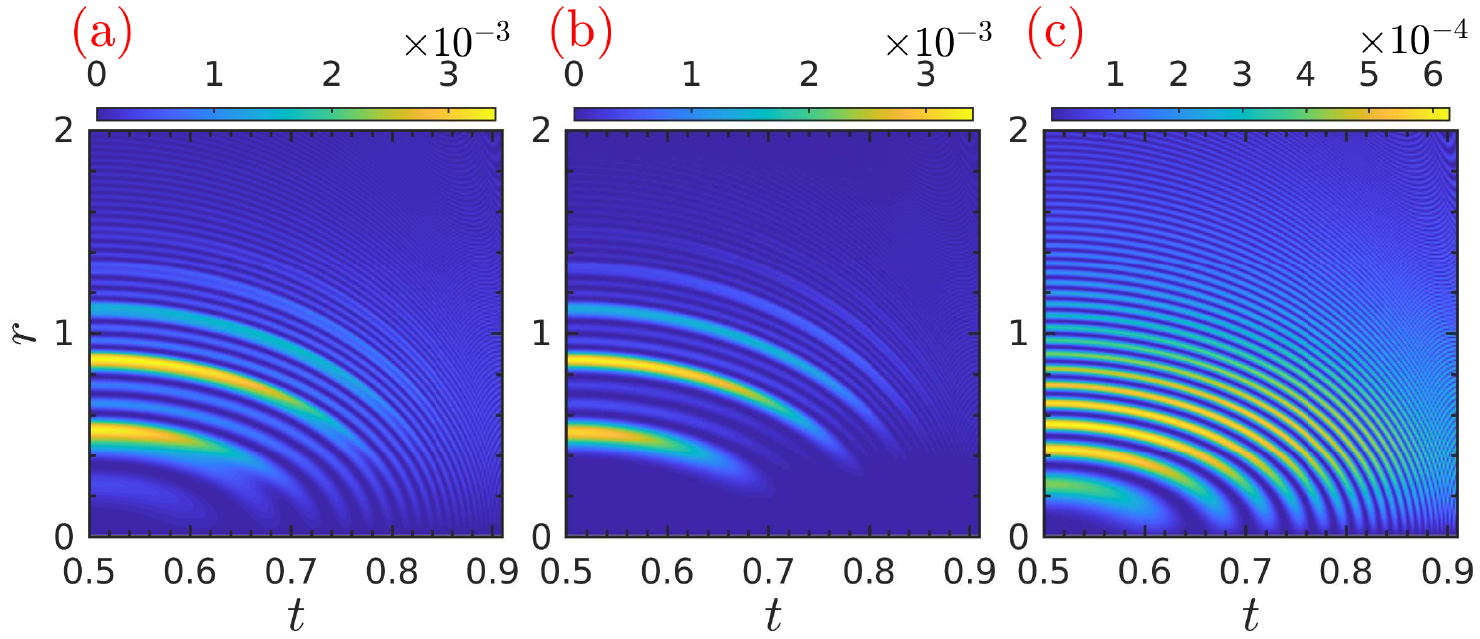}
    \caption{Double differential in transverse momentum parameter $r$ and lightfront momentum fraction $t$ of positron yield $d^2\tsf{P}/d t d r$. (a) all the harmonic channels open in the flat-top pulse; (b) only the harmonics accessible in a local monochromatic approach, $n\geq n_{*}$; (c) harmonics inaccessible to the local approach, $n<n_{*}$. The results are for $\xi=0.5$, $\eta=1$, $N=4$.}
    \label{fig:AngPlot1}
\end{figure}

As a further demonstration of the role of the extra harmonic channels, in~\figref{fig:AngPlot1}, we plot the combined lightfront and transverse momentum dependency of produced positron yield for the case $\xi=0.5$, $\eta=1$, $N=4$.
There are three main differences brought about by the pulse envelope:
i) since the suppression of larger lightfront momentum fractions is softened in the finite pulse cases, there are more harmonics visible at large/small $t$ part of the spectrum in~\figref{fig:AngPlot1}(a) compared to the plot of harmonics only accessible in the locally monochromatic approach in~\figref{fig:AngPlot1}(b).
The contribution from  harmonics inaccessible to a local approach, is shown  in \figref{fig:AngPlot1}(c).
ii) there appear subharmonics between the main harmonic lines in~\figref{fig:AngPlot1}(b), which corresponds to the sub-peaks in the regularised delta function in \eqnref{eqn:Inflat}. The sub-threshold harmonics lines in~\figref{fig:AngPlot1}(c) can further brighten these subharmonics in the whole yield in~\figref{fig:AngPlot1}(a) but also lead to a splitting of the main harmonics at large/small $t$;
iii) a peak occurs at low transverse momentum (the $r\to0$ region) due to the opening of the $|n|\leq1$ harmonic channels in the flat-top pulse. This link can be understood by considering the regularised delta function in $r^{2}$ in \eqnref{eqn:Inflat}. In the locally-monochromatic approach, the radial position squared of each harmonic is given by:
\bea
 r^{2}_{\infty}(\vphi) = 2n\eta t(1-t) - [1+\xi^{2}(\vphi)], \label{eqn:r1aa}
\eea
and $r^{2}(\vphi) = r^{2}_{\infty}(\vphi)$. The minimum perpendicular momentum is then at $r^{2} \approx \lceil n_{\ast} \rceil \eta/2 -1$. However, in the flat-top pulse, considering the additional width of each harmonic supplied by the pulse envelope, we see $r^{2}(\vphi)$ is approximately in the interval given by $r^{2}_{\infty}(\vphi) \pm \eta t(1-t)/N$. So we should expect that there is a signal in the transverse momentum distribution (TMD) due to the pulse envelope, at smaller transverse momenta than predicted by the local approach. Because this is to do with the width of each harmonic and contains the length scale $2N\lambda$ (substituting $n\to1$ and $\eta \to \eta \pm \eta/2N$ in \eqnref{eqn:r1aa} leads to the additional width supplied by the pulse envelope), it can be related to interference on the length scale of the pulse envelope. (A similar behaviour of the appearance of a linear peak appearing at small $r$ was noted for the  Compton scattering in a flat-top pulse \cite{King:2020hsk}.) (It was also recently observed that the enhancement of harmonics at lower values of $r$ can be understood by considering angular-momentum conservation in the absorption of laser photons producing the pair \cite{Kohlfurst:2018kxg}.)

\begin{figure}[h!!]
    \centering
    \includegraphics[width=8.5cm]{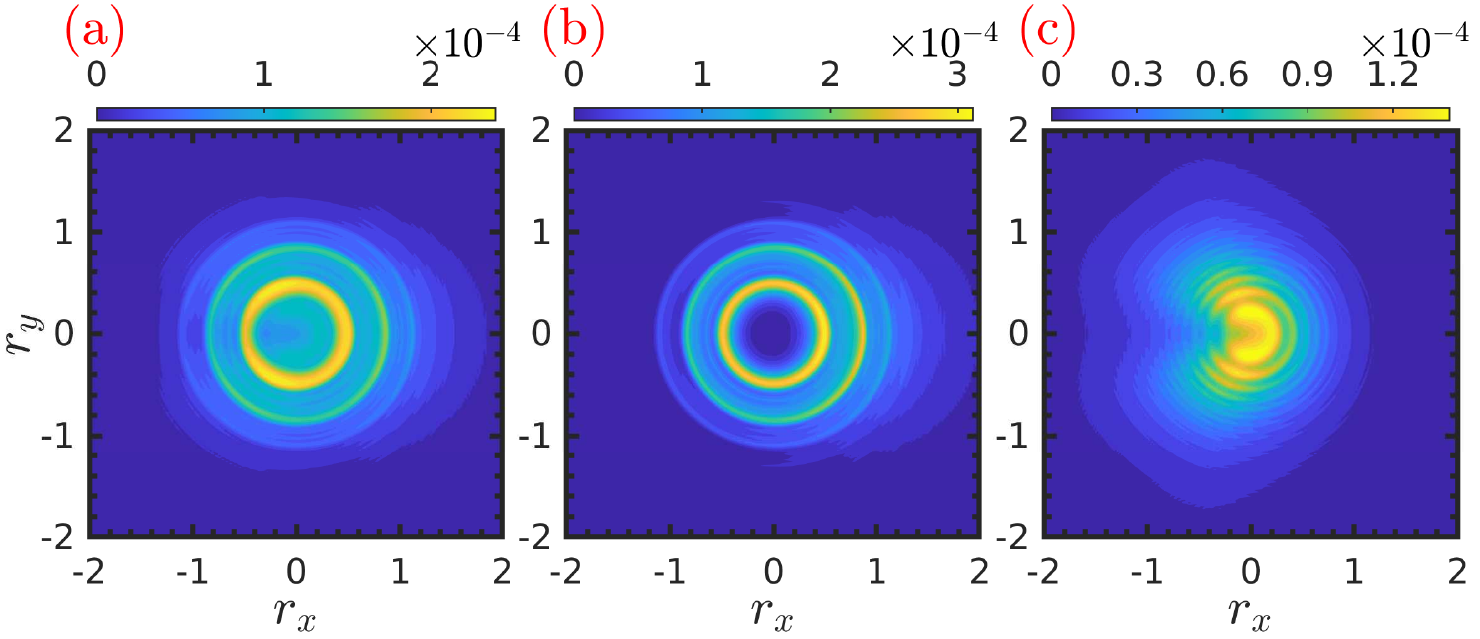}
    \caption{Double differential in the transverse momentum parameters of the positron yield $d^2\tsf{P}/d r_{x}d r_{y}$. (a) all the harmonic channels open in the flat-top pulse; (b) only the harmonics accessible in a local monochromatic approach, $n\geq n_{*}$; (c) harmonics inaccessible to the local approach, $n<n_{*}$. All the parameters are the same as in Fig.~\ref{fig:AngPlot1}.}
    \label{fig:asyPlot1}
\end{figure}

Finally, another role of the finite pulse duration shown in Fig.~\ref{fig:asyPlot1} is to induce an azimuthal asymmetry in the positron yield distribution~\cite{Titov:2015tdz}. As shown in the figure, this asymmetry appears not only in the distribution of the positron yield from the harmonics accessible in the local approach, but also in the distribution from the extra harmonic channels opened by the finite pulse effect (Fig. ~\ref{fig:asyPlot1}(b) and (c) respectively). This asymmetry affects the \emph{entire} TMD and is enhanced in \emph{short} pulses. Therefore, it is associated with the shape of the pulse and higher derivative effects. In contrast, in Sec.~\ref{Azimuthal}, we will demonstrate an asymmetry at the \emph{centre} of the TMD, to do with pulse-envelope interference.

\section{Linear Breit-Wheeler}
It was shown in the previous section that finite pulse envelope effects are particularly strong in the multi-photon regime at low background intensity. The reason for this is simply that the yield, when calculated with a local approach, is so strongly suppressed in this regime, that the effect due to the pulse envelope, which is otherwise small, becomes dominant. In this section, we analyse the contribution to the yield to leading order in $\xi^{2}$ - also known as \emph{linear} Breit-Wheeler. The dominance of pulse effects in the low-intensity (and strongly-suppressed) region has already been highlighted in the literature (see e.g. \cite{PhysRevLett.108.240406,seipt12b}), and so here, we concentrate on the role of the negative harmonics, and in doing so, will reveal a connection between photon polarisation channels and pulse duration.
\newline

We can acquire the leading-order perturbative probability, $\tsf{P}_{\ell}$ either by perturbatively expanding the integrand in \eqnref{eqn:P1} and keeping only $O(\xi^{2})$ terms, or by calculating the linear Breit Wheeler process from first principles in a plane wave background. In order to aid interpretation, we define a momentum contributed by the background field, $\nu \vkap$, through the relation:
\[
\nu\vkap+k = p+q,
\]
where $\nu = \bar{s}/2\eta$, which is the centre-of-mass energy squared in units of $2\eta m^{2}$. This is useful because $\nu$ is the ratio of lightfront momentum supplied by the background over the lightfront momentum supplied by the carrier wave (i.e. the monochromatic limit), and so quantifies how wide the pulse's bandwidth must be to facilitate the process. The probability for linear Breit Wheeler probability is then:
\be
\tsf{P}_{\ell} =\frac{\alpha}{\pi\eta} \int^{1}_{0} dt \int_{\nu_{\ast}}^{\infty} d\nu ~\frac{|\mbf{\tilde{a}}(\nu)|^{2}}{m^{2}}\,\left[\frac{1}{2}h(t) + \frac{\nu/\nu_{\ast}-1}{\nu^{2}/\nu^{2}_{\ast}}\right] \label{eqn:pert1}
\ee
where $\nu_{\ast}(t) = 1/[2\eta t(1-t)]$ is the threshold parameter required to create a pair with zero transverse momentum ($r^{2}=0$) and $\mbf{\tilde{a}}(\nu) = \int d\vphi\, \mbf{a}(\vphi)\,\mbox{e}^{i\vphi \nu}$.

The result in \eqnref{eqn:pert1} clearly demonstrates the threshold requirement $\nu>\nu_{\ast}$ to create a pair. We see that $\nu_{\ast} \propto 1/\eta$, i.e. the lower the photon energy, the higher the momentum required from the background to create a pair. For a plane-wave pulse then, the contribution depends on how the Fourier spectrum decays as the frequency is varied away from the carrier frequency. For the flat-top potential,
\bea
\frac{|\mbf{\tilde{a}}(\nu)|^{2}}{m^{2}} =\frac{\xi^{2}\Phi^{2}}{2}
\left[\sinc^{2}\frac{\Phi(\nu+1)}{2} +\sinc^{2}\frac{\Phi(\nu-1)}{2}\right]
\label{eqn:fourierPert}
\eea
i.e. the spectrum has peaks at $\nu = \pm 1$, but each with a width that decays as $\sim 1/\Phi^{2}$. Here we see the reason why the pulse envelope effect is dominant in the multiphoton regime. For the monochromatic case, $\tsf{P} \sim \xi^{2n_{\ast}}$, with $n_{\ast}=2(1+\xi^{2})/\eta$, and so when $\eta \ll 1$, which is typical for modern-day laser-particle experiments, there is a strong decay in the pair yield as $\xi$ is reduced since $n_{\ast}\gg1$. Contrast this with the linear Breit-Wheeler contribution from the envelope, which only decays as $\xi^{2}$, but is suppressed by a constant factor that depends on the envelope's bandwidth. Eventually, for sufficiently small $\xi$, the linear Breit-Wheeler effect, which originates from the pulse envelope, will dominate the total signal for pair creation

\begin{figure}[h!!]
    \centering
    \includegraphics[width=6cm]{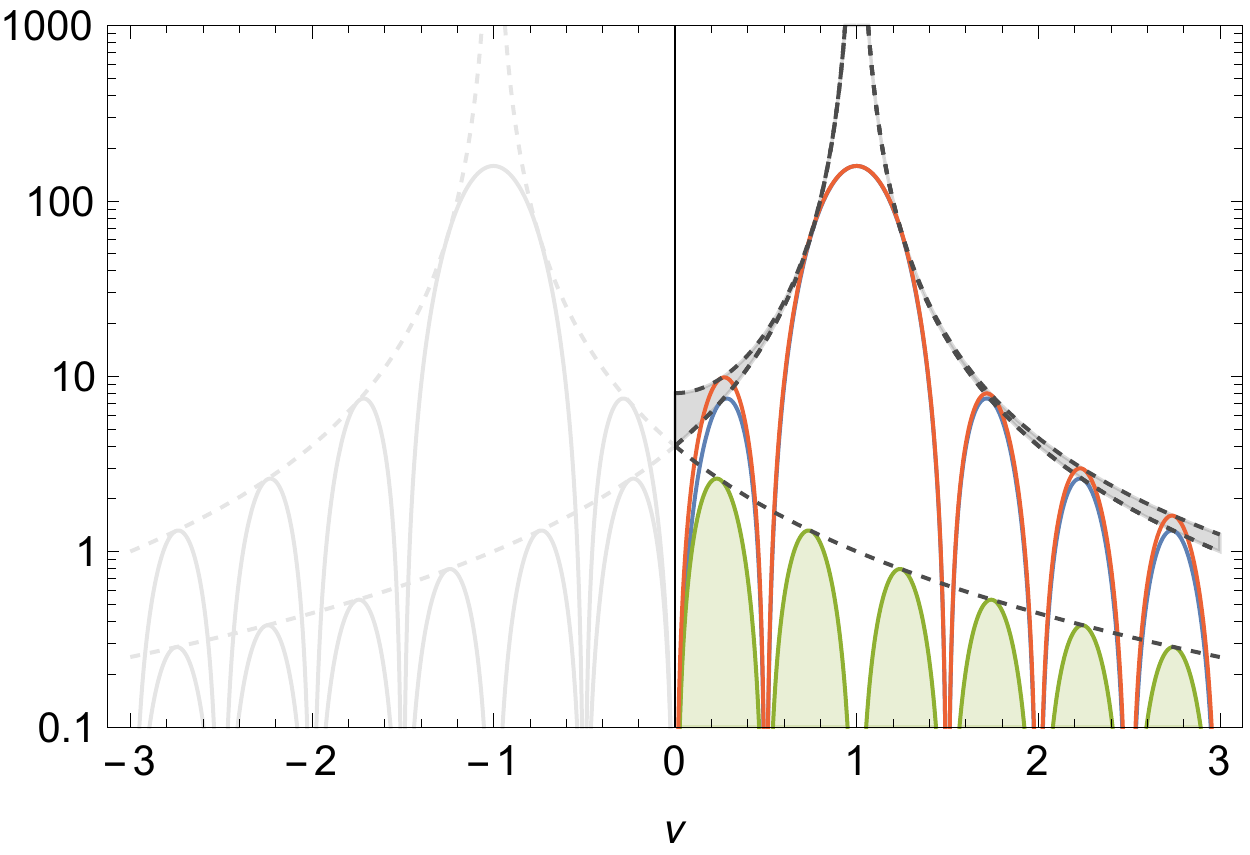}
    \caption{Fourier spectrum of potential for a two-cycle flat-top pulse ($\Phi=4\pi$). Only the absorption of positive frequencies can kinematically contribute to pair-creation, but in a short pulse, these can arise from the wide bandwidth of the \emph{negative} frequency part of the spectrum. Envelopes are plotted with dashed lines and the contribution from the $n=-1$ harmonic, and the net difference, are both highlighted.}
    \label{fig:FTspec}
\end{figure}

The role of negative harmonics can also be clearly seen in the perturbative result. In~\figref{fig:FTspec}, we plot the Fourier transform of the flat-top pulse spectrum, \eqnref{eqn:fourierPert}. The spectrum decays away from the positive [negative] frequency peak, approximately as $\sim(\nu-1)^{-2}$ [$(\nu+1)^{-2}$]. The postive (negative) frequency peaks correspond to the $n=1$ ($n=-1$) harmonics of the background. For a sufficiently short pulse, equivalently, broad bandwidth, there is a significant contribution from the negative frequency peak, that can extend to the positive frequency range.
This is another visualisation of the effect discussed in the Mandelstam plot \figref{fig:mandelstam1}: negative harmonics can contribute in kinematically-accessible regions. We emphasise that here in the perturbative case, $n=-1$ corresponds to the negative frequency component of the pulse. One should contrast this with the non-perturbative case discussed in the previous section and for larger values of $\xi$, where $n=-1$ corresponds to one \emph{net} `photon' being emitted back to the background.

\subsection{Polarisation signal}

The contribution of the negative harmonic to linear Breit-Wheeler suggests a further effect due to the pulse envelope: on the relative contribution of each photon-polarisation channel. This can be seen by considering a head-on collision between the photon and a plane-wave laser pulse. The potential for a circularly-polarised pulse can be written in the form:
\bea
a(\vphi) = g(\vphi)\left[\Lambda_{-}\mbox{e}^{i\vphi}+\Lambda_{+}\mbox{e}^{-i\vphi}\right], \label{eqn:alambda}
\eea
where $g(\vphi)$ is the envelope and $\Lambda_{\pm} = (\Lambda_{1}\pm i\Lambda_{2})/2$ are lightfront helicity states of the probe photon:
\bea
\Lambda_{1,2} = \eps_{1,2} - \frac{k\cdot \eps_{1,2}}{k\cdot \vkap}\vkap,
\eea
for a head-on collision with $k=k^{0}(1,0,0,-1)$, and in the co-ordinates used so far in this paper, $\eps_{1} = (0,1,0,0)$, $\eps_{2} = (0,0,1,0)$. From~(\ref{eqn:alambda}), we see that different helicity states are associated with the positive and negative frequency parts of the background. Whereas a circularly-polarised probe photon will couple to just the $\Lambda_{+}$ or $\Lambda_{-}$ state alone, a linearly-polarised photon, being a superposition of circularly-polarised states, will be able to couple the $\Lambda_{+}$ and $\Lambda_{-}$ states together at the level of the probability (which contains a $(a\cdot \epsilon)^{2}$ term, for photon polarisation $\epsilon$). Therefore, for a linearly-polarised photon, a part of the probability will be proportional to the overlap of the positive and negative frequency parts of the spectrum. In this way, a connection between pulse duration and probe photon polarisation is made.

For a linearly polarised photon propagating in the circularly polarised background, the linear Breit-Wheeler probability is
\begin{align}
\tsf{P}_{\ell}&=\frac{\alpha}{\pi\eta}\int^{1}_{0} \ud t\int^{\infty}_{\nu_{*}} \ud\nu \left[\frac{|\tilde{\bm{a}}(\nu)|^2}{m^2}\left(\frac{1}{2}h(t)+\frac{\nu/\nu_{*}-1}{\nu^{2}/\nu^{2}_{*}}\right) \right. \nn \\
       & \left. \qquad\qquad\qquad\qquad~~+ \Gamma \frac{|\tilde{a}_{y}(\nu)|^{2} -|\tilde{a}_{x}(\nu)|^{2}}{2m^{2}\nu^{2}/\nu^{2}_{*}}\right]\,,
\end{align}
where $\tilde{a}_{x}$ ($\tilde{a}_{y}$) is the $x$ ($y$) component of the Fourier transformed vector potential, $\Gamma=1$ if the photon is in polarisation state $\Lambda_{1}$, and $\Gamma=-1$ if the photon is in polarisation state $\Lambda_{2}$. Defining $\tilde{g}(\nu)=\int d\vphi\, g(\vphi)\exp[i\vphi\nu]$, we see that the unpolarised part of the probability is proportional to:
\[
|\tilde{\bm{a}}(\nu)|^2 \propto |\tilde{g}(\nu+1)|^{2}+|\tilde{g}(\nu-1)|^{2}.
\]
However, the polarised part is proportional to the overlap of
the positive and negative frequency components:
\[
|\tilde{a}_{y}(\nu)|^{2} -|\tilde{a}_{x}(\nu)|^{2} \propto \tilde{g}(\nu+1)\tilde{g}^{\ast}(\nu-1) + \trm{c.c.}
\]
This overlap of the functions is an expression of the interference in Fourier space, of the contribution from positive and negative frequency components. Therefore this coupling of polarisation channels to pulse shape is a higher derivative effect. To measure the relative importance of the photon polarisation, we define the quantity
\bea
\tsf{R}=\frac{|\tsf{P}_{\ell}(\Gamma=-1)-\tsf{P}_{\ell}(\Gamma=+1)|}{\tsf{P}_{\ell}(\Gamma=-1)+\tsf{P}_{\ell}(\Gamma=+1)}. \label{eqn:Rdef}
\eea

\begin{figure}[t!!]
    \centering
    \includegraphics[width=8.7cm]{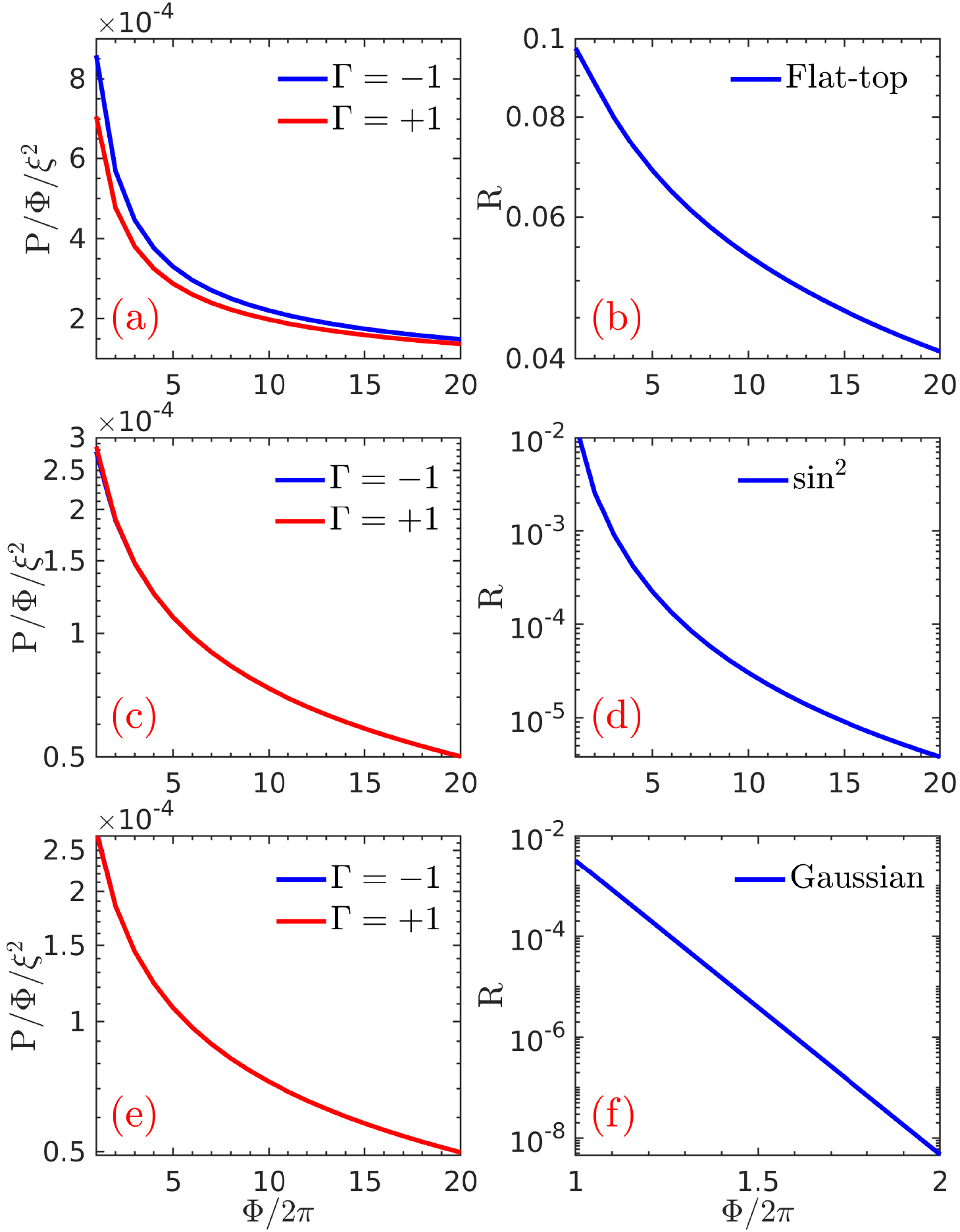}
    \caption{Total yield, for $\eta=2$, of the positrons from the linear Breit-Wheeler process normalised by the pulse duration $\Phi$ and intensity $\xi$ (left column) and the relative importance of the probe photon's polarisation in the positron yield $R$, given in \eqnref{eqn:Rdef} (right column). First row: flat-top pulse, $g(\vphi)=1$ if $0<\vphi<\Phi$, otherwise $g(\vphi)=0$; Middle row: sine-squared pulse, $g(\vphi)=\sin^{2}[\vphi/(2N)]$ if $0<\vphi<\Phi$, otherwise $g(\vphi)=0$; Last row: Gaussian pulse, $g(\vphi)=\exp[-2.3\vphi^{2}/(N\pi)^{2}]$. The background potential is circularly polarised and the probe photon is linearly polarised. The pulse phase duration is defined as $\Phi=2N\pi$.}
    \label{fig:linear}
\end{figure}

The linear Breit Wheeler positron yield for different probe photon polarisations, and how it is affected by pulse duration, is illustrated in Fig.~\ref{fig:linear}. In general, the yield increases with pulse duration, but it does this slower than linearly with $\Phi$ (unlike a local approach, which increases the yield linearly with $\Phi$). This is because the bandwidth decreases and so too, the number of photons from the background that would correspond to a centre-of-mass energy above the threshold for making a pair. This behaviour depends on the shape of the pulse: slightly different dependencies are shown in Fig.~\ref{fig:linear}(a) (flat-top pulse),  Fig.~\ref{fig:linear}(c) (sine-squared pulse) and Fig.~\ref{fig:linear}(e) (Gaussian pulse). The effect of the pulse duration on the ratio of polarised probabilities is shown in the second column of Fig.~\ref{fig:linear}. Similar to the yield, the ratio decreases as the pulse duration is increased, and this is due to reduction of the overlap of positive and negative frequencies of the background in the pulse spectrum. For a sine-square pulse, this decrease is rather slow: for a single-cycle pulse the ratio is 10\%, falling to 4\% for a $20$-cycle pulse. For a sine-squared and Gaussian pulse (Fig.~\ref{fig:linear}(d) and (f) respectively), the decrease is much stronger, showing the importance of the pulse shape in this effect.

\section{Azimuthal asymmetry}~\label{Azimuthal}
In this section, we calculate a signal of azimuthal asymmetry at the centre of the positron TMD, even though the plane-wave pulse is circularly polarised. The asymmetry for different pulse shapes is compared.

\begin{figure}[t!!]
    \centering
    \includegraphics[width=8.7cm]{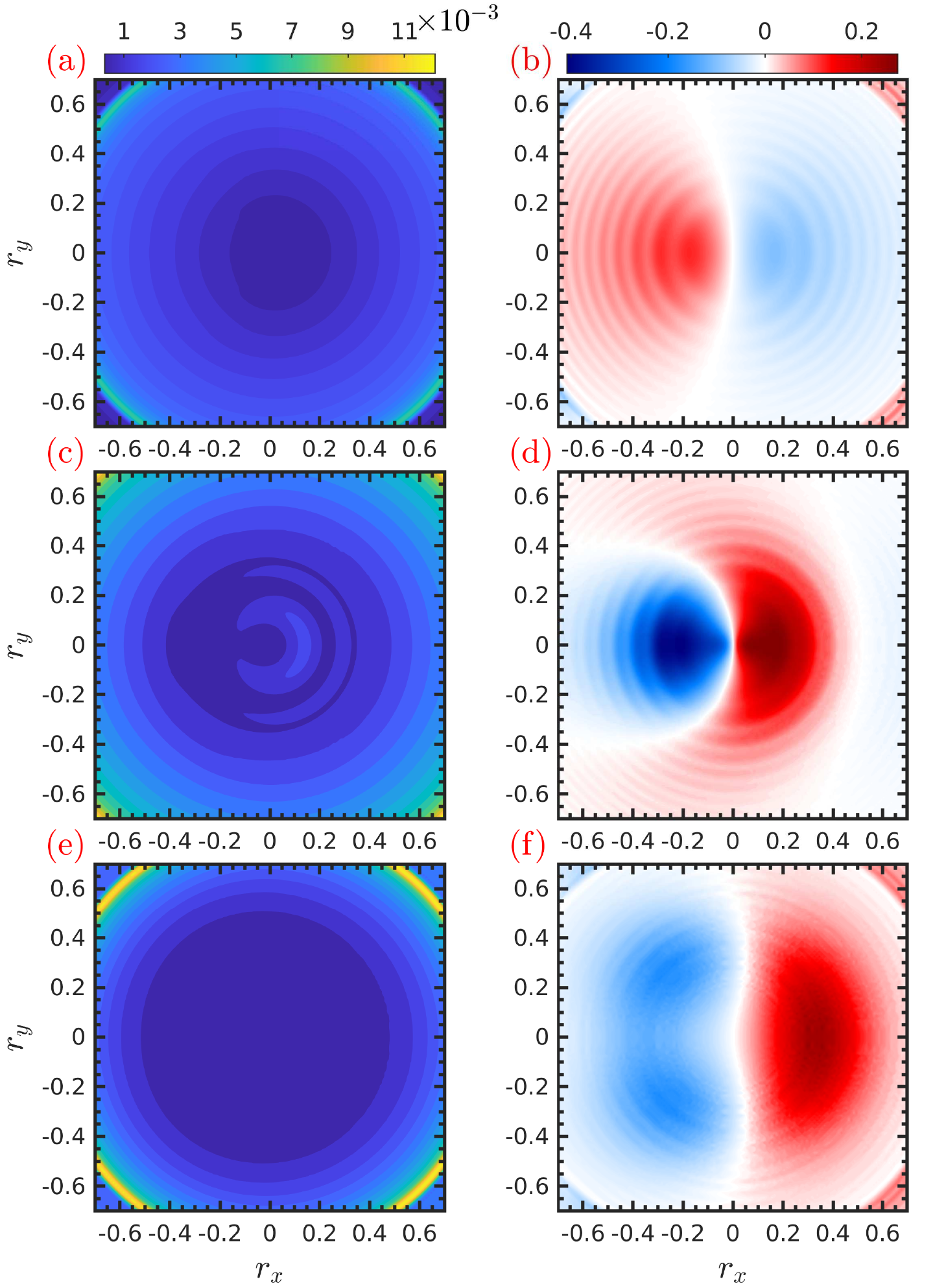}
    \caption{Azimuthal asymmetry in the transverse momentum distribution of the positron produced in a flat-top pulse for $\eta=2$, $N=16$. Left column: plot of the angular spectrum, $d^{2}\tsf{P}/dr_{x}dr_{y}$. Right column: asymmetry measure, $\mathcal{A}(r_x,r_y)$. Upper panels, $\xi=0.5$; Middle panels, $\xi=1.0$; Bottom panels, $\xi=1.5$.}
    \label{fig:azy1}
\end{figure}

The asymmetry induced by the pulse envelope can be illustrated by formulating the relative difference of the transverse spectrum and the azimuthal-averaged spectrum:
\[
\mathcal{A} = \frac{d^{2}\tsf{P}/dr\,d\psi}{\frac{1}{2\pi}\int_{0}^{2\pi}\left[d^{2}\tsf{P}/dr\,d\psi\right]\,d\psi}-1.
\]
The result for a flat-top background with $\xi=0.5$, $\eta=2$ and $N=16$, in \figref{fig:azy1}, (using $r_{x} = r \cos \psi$ and $r_{y} = r\sin \psi$), amounts to uncovering a dipole-like distribution, which is symptomatic of having linear polarisation. This dipole structure is strongest at transverse momenta $r$, which correspond to contributions from the pulse envelope for harmonics below the locally-monochromatic threshold.  Whereas the carrier frequency is circularly polarised, the pulse envelope multiplies both polarisation components of the potential $a$ with the same factor, and hence acts as a `linearly polarised' background. We can demonstrate this by writing the Fourier transform as:
\[
\tilde{a}(\nu) = m\xi\left[\tilde{g}(1+\nu)\Lambda_{-}+\tilde{g}(-1+\nu)\Lambda_{+}\right].
\]
In a monochromatic wave, $\tilde{g}(\cdot) \to \delta(\cdot)$ and each frequency component is associated with a different helicity state. However, if the pulse is finite in extent, and we pick $\nu =\nu_{\tiny\tsf{pulse}}\approx 1/2N$, to correspond to the pulse envelope frequency, for a symmetric spectrum (such as the flat-top, sine-squared and Gaussian examples), $\tilde{g}(1+\nu_{\tiny\tsf{pulse}}) \approx \tilde{g}(-1+\nu_{\tiny\tsf{pulse}})$, and hence:
\[
\tilde{a}(\nu_{\tiny\tsf{pulse}}) \approx m\xi\tilde{g}(1+\nu_{\tiny\tsf{pulse}}) \Lambda_{1},
\]
i.e. linearly polarised.

\begin{figure}[t!!]
    \centering
    \includegraphics[width=8.7cm]{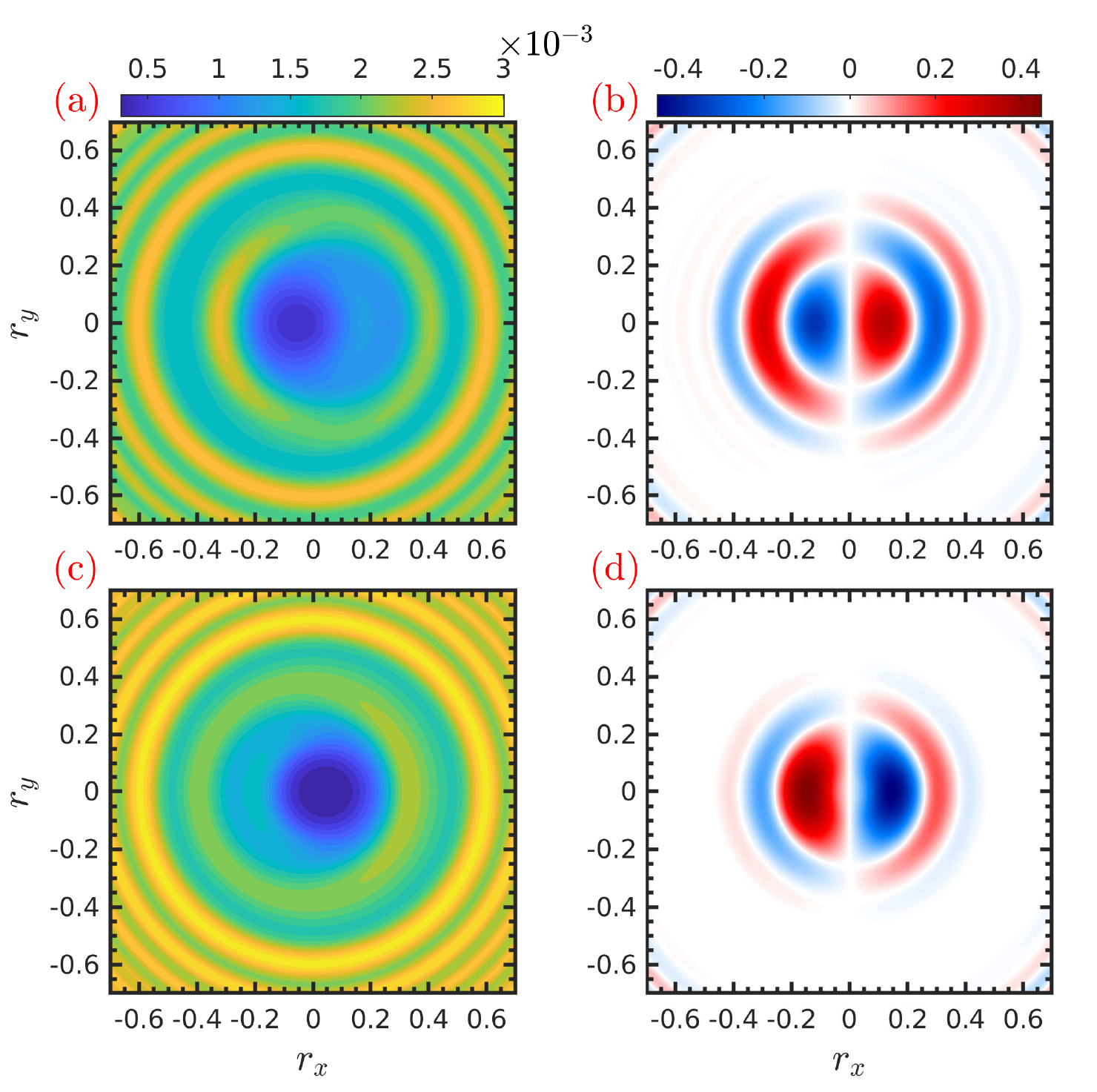}
    \caption{Azimuthal asymmetry in the transverse momentum distribution of the positrons. Left column: plot of the angular spectrum, $d^{2}\tsf{P}/dr_{x}dr_{y}$. Right column: asymmetry measure, $\mathcal{A}(r_{x},r_{y})$. Upper panels: sine-squared pulse, $g(\vphi)=\sin^{2}[\vphi/(2N)]$ if $0<\vphi<2N\pi$, otherwise $g(\vphi)=0$; Bottom panels: Gaussian pulse with the envelope $\exp[-2.3\phi^{2}/(N\pi)^{2}]$. The other parameter are given as $\xi=1$, $\eta=2$ and $N=16$.}
    \label{fig:azy2}
\end{figure}

The effect persists in pulses with smoother edges as illustrated in Fig.~\ref{fig:azy2}(a) and (b) for a sine-squared pulse and in Fig.~\ref{fig:azy2}(c) and (d) for a Gaussian pulse. We note that in these examples, $N=16$, i.e. this is \emph{not} the short-pulse effect described in \figref{fig:AngPlot1}.
We also note the interference fringes (although not their magnitude) depend sensitively on the shape of pulse envelope; interference peaks appear at different positions for different envelopes, and there appear more interference fringes because of the variation of the local intensity $\xi(\vphi)$.  A noteworthy feature of the asymmetry at the centre of the positron TMD is that it becomes much weaker when $\xi$ is lowered below $\xi=1$ for the sine-squared and Gaussian pulses, suggesting an all-order interaction  (but persists for $\xi < 1$ in the special case of the  flat-top pulse).

\section{Discussion}
The nonlinear Breit-Wheeler process is often calculated using some local approximation, such as the locally constant field approximation (LCFA), or the locally monochromatic approximation (LMA). The focus of this paper has been to identify signals of strong-field QED in the nonlinear Breit-Wheeler process, that are beyond local approaches. In order to achieve this, we calculated the process in a toy-model flat-top pulse potential, which differs only from the infinite monochromatic background in that it has edges. The advantage of this background, is that one can make analytical progress and cast the form of the probability for this finite pulse, in a similar form to the probability in an infinite monochromatic wave, and study the differences. Any difference is then due to the finitude of the pulse, and not due to variations of the field strength across the pulse (which, e.g. the locally monochromatic approximation would partially capture). We also calculated some spectra in smoothly varying pulses, with sine-squared and Gaussian envelopes, to investigate how the shape of the pulse envelope influences `beyond local' signatures.
\newline

What is missed in a local monochromatic (and locally constant) approach, can be seen by considering the two steps made in approximating the Kibble mass \cite{kibble64}, $\mu$, where:
\[
\mu = 1 + \langle \mbf{a}^{2} \rangle - \langle\mbf{a}\rangle^{2}; \quad \langle f \rangle = \theta^{-1}\int_{\vphi-\theta/2}^{\vphi+\theta/2} f(x)dx.
\]
Suppose we consider $\langle \mbf{a} \rangle$ as an example, and apply the LMA to the function $\mbf{a}(\vphi) = \mbf{e}_{x} g(\vphi/\Phi) \cos\vphi$. First, we can write:
\[
\langle \mbf{a} \rangle = \frac{\mbf{e}_{x}}{\theta}\left[F\left(\frac{x}{\Phi}\right)\sin x+G\left(\frac{x}{\Phi}\right)\cos x\right]^{\vphi+\theta/2}_{\vphi-\theta/2},
\]
where by repeated integration by parts, one finds:
\bea
F\left(\frac{x}{\Phi}\right) &=& g\left(\frac{x}{\Phi}\right) - \frac{1}{\Phi^{2}}g''\left(\frac{x}{\Phi}\right) + \ldots \nn \\
G\left(\frac{x}{\Phi}\right) &=& g'\left(\frac{x}{\Phi}\right) - \frac{1}{\Phi^{3}}g'''\left(\frac{x}{\Phi}\right) + \ldots \nn
\eea
For smooth, well-behaved envelopes, so far no approximation has been made. The LMA then consists of two approximations: i) higher derivatives of $g$ are discarded; ii) the envelope is locally expanded, keeping only the leading order term i.e. $g[(\vphi\pm \theta/2)/\Phi] \approx g(\vphi/\Phi)$. The first approximation (also referred to as the slowly-varying-envelope approximation) corresponds to neglecting derivatives of the envelope, and the second to neglecting pulse-envelope interference (which would also include some derivatives of the envelope). Therefore, even if the pulse is long, if the rising and falling edges are steep enough, then signatures beyond a local approximation, can still persist. As a result, we have identified effects on particle spectra due to: i) pulse-envelope interference; ii) pulse shape effects (contributions from higher derivatives of the rising and falling edge of the pulse).

Other works have already highlighted some interference effects for very short (few cycle) pulses; the added interest here, was to show that there are other pulse-length interference effects that persist even in long (many-cycle) pulses, the magnitude of which does not depend strongly on the pulse shape. One may think that, if the pulse is made long enough, outgoing particle spectra should eventually tend to those in a monochromatic wave. However, the key difference in the finite pulse case, is that the probe particle (in this case, a photon), is, at some point, outside the pulse, and must enter (and exit) it. In the infinite monochromatic case, the probe particle is \emph{always} inside the background. The act of entering (and exiting) the pulse, is associated with the probe particle experiencing field gradients. The longer the pulse, the lower the gradients, but the longer the scale on which they are probed.
\newline

For nonlinear Breit-Wheeler, pulse envelope interference affected the positron transverse momentum distribution by providing an extra `pulse envelope peak' at smaller transverse momenta than the threshold for a locally monochromatic approach. The position of this pulse envelope peak is associated with the long wavelength of the pulse envelope. The peak was found to lead to a linearly-polarised signal in a flat-top, sine-squared and Gaussian pulse, the magnitude of which did not depend significantly on pulse shape. In contrast, effects were also identified that are related to the pulse shape and the contribution from higher derivatives of the envelope. In a flat-top pulse, the relative importance of Breit-Wheeler for different photon polarisations was seen to be linked to the pulse duration, but was suppressed in the smoother sine-squared and Gaussian backgrounds. The short-pulse asymmetry across the entire positron transverse momentum is another effect sensitive to the pulse shape, and was demonstrated for the flat-top background. A significant widening of the lightfront momentum spectrum was also found in a flat-top background, which was independent of the pulse duration.

To conclude: if pulse shaping methods could be used to generate steeper pulse edges, envelope shape effects can be enhanced, and if low transverse pair momenta could be detected, pulse-length interference effects could be measured in experiment.

\section{Acknowledgments}
BK acknowledges support from the Engineering and Physical Sciences Research Council (EPSRC), Grant No. EP/S010319/1.
S. Tang acknowledges support from the Young Talents Project at Ocean University of China and the National Natural Science Foundation of China, Grants No.12104428.
\appendix

\bibliography{current}

\end{document}